# Biobeam – Rigorous wave-optical simulations of light-sheet microscopy

Martin Weigert[1,2], Eugene W. Myers[1,2,3] & Moritz Kreysing[1,2]

[1]Max Planck Institute of Molecular Cell Biology and Genetics, Dresden, Germany

[2]Center for Systems Biology Dresden, Germany

[3]Faculty of Computer Science, Technische Universität Dresden, Germany

**Sample-induced image-degradation remains an intricate wave-optical problem in light-sheet microscopy. Here we present *biobeam*, an open-source software package that enables to simulate operational light-sheet microscopes by combining data from $10^5$–$10^6$ multiplexed and GPU-accelerated point-spread-function calculations. The wave-optical nature of these simulations leads to the faithful reproduction of spatially varying aberrations, diffraction artifacts, geometric image distortions, adaptive optics, and emergent wave optical phenomena, and renders image-formation in light-sheet microscopy computationally tractable.**

Light-sheet fluorescence microscopy is a popular tool for the volumetric imaging of developing organisms [1–5]. As light-sheet microscopes continue to be developed for progressively bigger biological samples [6, 7], there is an increasing need for computers to process gigantic sets of imaging data and extract biologically relevant information [8]. With sample size, however, also light-scattering induced imaging artifacts become increasingly prevalent. These are currently dealt with on a mostly phenomenological basis. A faithful forward model of the wave-optical imaging process, however, would *i)* enable rigorous benchmarks of deconvolution and segmentation strategies against ground-truth data [9], *ii)* serve as training platforms for machine learning approaches for image restoration and information extraction and *iii)* leverage the efficient use of adaptive optics [10, 11] to prevent sample-induced image degradation during the acquisition process.

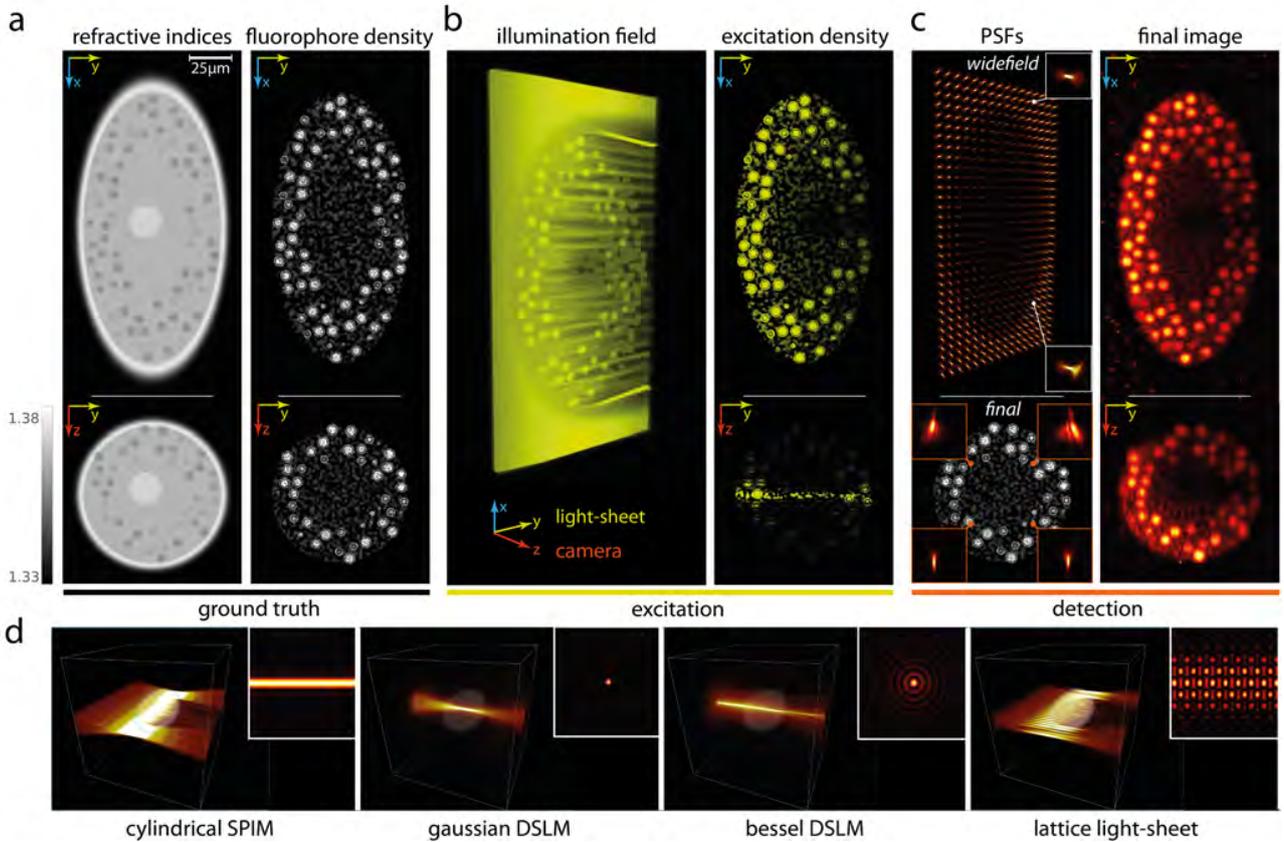

**Figure 1:** *Rigorous wave-optical simulation of image formation process in light-sheet microscopy*
(a) Synthetic tissue phantom of a multicellular organism ($100 \times 200 \times 100 \mu m$) comprising a complex refractive index distribution (left, $n = 1.33 - 1.38$) and a fluorophore distribution of interest (right). (b) Wave optical simulation of the illuminating light sheet and resulting excitation distribution within the sample at a given z position. (c) Partially coherent simulation of the detection path by multiplexed calculation of all independent point spread functions (left) and the resulting simulated camera image combining illumination and fluorescence path of light through the scattering sample. (d) Alternative light-sheet modalities (see also Supplementary Movie 1 & 5 and main text for details).

Predicting light-tissue interactions is particularly demanding when leaving the single scattering regime [12] or strictly diffusive transport [13], and despite significant computational advances [14] generally applicable solutions remain prohibitively computationally costly [15] to simulate mi-croscopes. Even when constraining simulations to the biological relevant case of predominantly forward scattering tissue [10, 16], individual point spread function (PSF) calculations still require multiple seconds [17]. As aberrations deep inside tissues are unique for virtually each point in the sample [10, 12] a realistically large biological specimen, i.e. an embryo with a volume $\sim 100 \mu m^3$ would require $10^5$–$10^6$ volumetric PSF calculations in order to faithfully mimic the wave-optical imaging process which with current methods would take several weeks. As a consequence, attempts to simulate microscopic imaging have been limited to ray optics [18] or convolution with a constant PSF [19], approaches that do not reflect the wave-optical nature of light interaction with optically heterogeneous biological samples. Here, we report on *biobeam*, a software package that enables the first rigorous simulations of wave-optical image formation in light-sheet microscopes by *i)* a novel multiplexing scheme for PSF calculations and *ii)* efficient GPU parallelization. The pipeline underlying our software builds on the observation that the beam-propagation model (BPM) for fiber optics [20, 21] can also be used to mimic scattering biological cells [22]. To guarantee good accuracy of BPM beyond strictly paraxial wave propagation, we use the exact propagator together with a locally adapted expansion of refractive indices (see Supp. Notes 1 and 4 for details and validation against analytically tractable scattering models). To massively reduce the computational cost for $\sim 10^6$ wave-optical PSF calculations, we introduce a novel multiplexing scheme. In this we exploit the fact that in typical imaging scenarios, a single camera image can be constructed from sets of 100s to 1000s of mutually independent, spatially varying, non-overlapping PSFs. Such sets can be calculated within single, highly multiplexed simulations (see Supp. Fig. 3 and Supp. Notes 8), similar to the operational principle of spinning highly multiplexed confocal measurements of spinning disc microscopes [23, 24]. Our software is further accelerated $\sim 20$ fold by the efficient use of GPU implementations for all low-level calculations. In this way, the propagation of an arbitrary light field such as the multiplexed set of $\sim 1000$ PSFs through a typical volume of $1024^3$ voxels takes less than 500ms on a single graphics card ( cf. Supp. Notes 5, Table 2, Supp. Video 4), which constitutes a 20.000 fold acceleration compared to the sequential calculation of PSFs.

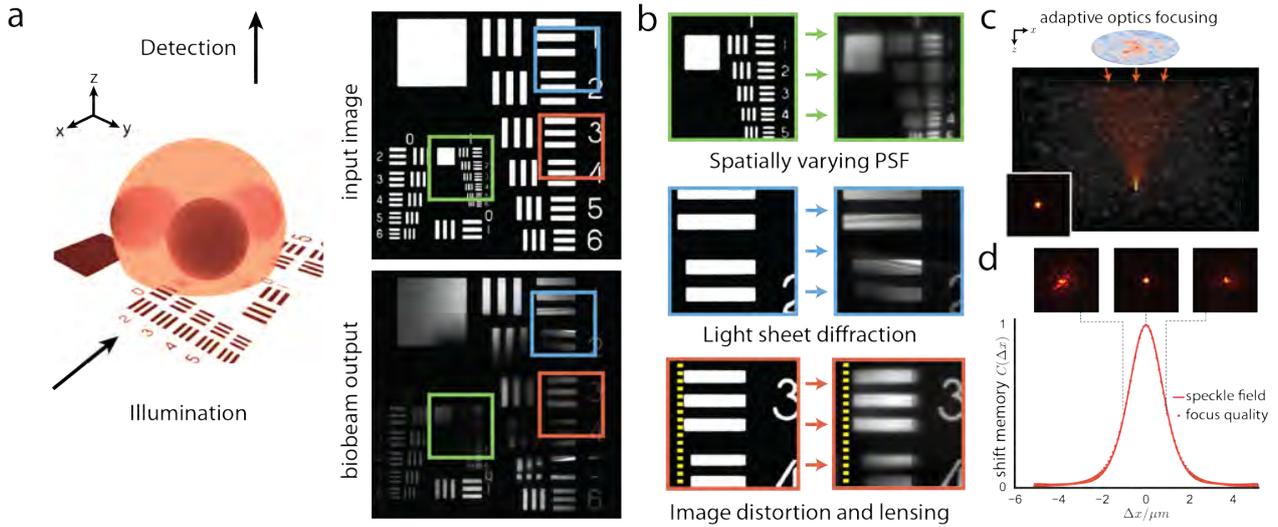

**Figure 2:** *Optical capabilities of the biobeam image-formation pipeline*
(a) Test chart at the mid-section of an optically heterogeneous embryo-model ($n = 1.35 - 1.39$ diameter $140\mu m$), is illuminated by cylindrical light sheet ($NA = 0.15$), and imaged from an orthogonal position ($NA = 0.6$). (b) Details of these wave-optically calculated images reveal *i)* spatially varying image blur, contrast loss and absorption induced by the heterogeneity of the sample, *ii)* diffraction artifacts from the light-sheet-typical coherent illumination, and *iii)* even geometric image distortions as lensing and split-screen type image distortionsobject displacements lensing. (c) *Biobeam* is further capable of adaptive optics simulations by which reversal of guide star emitted light fields yields perfect foci in scattering tissues. (d) Adaptive optics simulations faithfully reproduces the shift-shift memory effect, an emergent wave-optical phenomenon, here at 4 mean-free-paths deep inside the tissue.

We first demonstrate the power of our software package, by presenting the first wave-optical simulation of the volumetric image-formation process in a light-sheet microscope [5, 9, 18]. Here, a cylindrical sheet of light is propagated at a specific axial position through a given refractive and fluorescently labeled embryo model (Fig. 1a and Supp. Video 1 & 7). The fluorescence excitation at every point in the volume is calculated and a full set of PSFs of the detection system at the focal plane is obtained by propagating light from multiplexed diffraction-limited point-sources orthogonally through an idealized, refocusing lens towards the camera (Fig. 1c, upper left, Supp. Video 2). By this we obtain a quasi-continuum of spatially-dependent, volumetric PSFs with position-dependent aberrations stemming from sample-induced distortions induced on both the illumination and the detection path (Fig. 1c left, see also Supp. Fig. 4 for calculated aberration maps). Convolving the exhaustively sampled set of spatially-varying PSFs with the fluorescent object finally yields the wave-optical image as seen by the camera (Fig. 1c, right). These simulations are particularly demanding because of the large grid size (i.e. $1024 \times 2048 \times 1024$ voxels for Fig. 1) and sampling density requirements (typically $0.5-5\mu m$), here resulting in a total of $10^6$ PSF calculated in well under 10 minutes on a single graphics card (see also Supp. Videos 1 & 2), whereas the non-multiplexed calculations would have required more than two months on a single CPU. As a result of this computational pipeline *biobeam* generates faithfully calculated 3D microscopy data sets that account for both refraction and diffraction based imaging artifacts. These include image blur and contrast loss, spatially varying PSFs, speckle artifacts, image granularity, as well as sample-induced geometric distortions such as lensing, image displacements and split-screen type double images (cf. Fig. 2a/b, Suppl. Video 3, Suppl. Fig. 1 for benchmarks against analytically tractable models). As a particularly flexible software package, *biobeam* provides further pre-implemented illumination modalities, including Bessel lattices (Fig. 1d, right, Supp. Video 5), and allows the practical extraction of sample induced aberrations as spatially resolved Zernike maps (see Supp. Fig. 4 and Notes 9).

Next, we demonstrate *biobeam*'s capability to accurately simulate wave-optical effects relevant to adaptive optics imaging [10]. It is well-established and exploited in imaging that perfect imaging foci can be created behind totally opaque screens [25] and within biological samples by appropriately shaping the wave front before entry into the scattering medium. We recapitulated this finding by explicit simulations of light propagating through a spatially extended synthetic tissue sample consisting of Perlin-noise with refractive index $n = 1.36 \pm 0.03$). As expected, we find also in simulations that conjugating the wavefront point-like guide stars at the sample surface allows to generated diffraction-limited foci inside these scattering tissues (Fig. 2c, Supp. Video 6). Further we show that *biobeam* is capable of reproducing the shift-shift memory effect, an emergent wave optical phenomenon responsible for significant robustness of adaptive imaging against lateral focus displacements. In agreement with recent analytical arguments and their empirical confirmation [10], we find by our computational microscopy experiments that the average persistence of the laterally shifted focus is precisely limited by the autocorrelation length of a speckle pattern that would result from an incident plane wave. This phenomenon is accurately reproduced even 4 mean free path length inside the tissue (Fig. 2d). While currently using a variant of the beam propagation method, the here introduced multiplexing strategy of PSF calculations is principle also compatible with other low-level field stepping algorithms, including those that iteratively account for multiple backreflections [14, 15] where more precision seems appropriate.

We conclude that *biobeam* enables the wave-optically faithful simulation of light-sheet microscopes. This renders the biological imaging a process fully tractable by computers, thus providing the link between wave-optically recorded image material, and the ground truth object. Prospectively, we see *biobeam* helping to improve microscope design, enhancing deconvolution and segmentation strategies by providing realistic imaging data-sets along with ground truth data, and paving the way for a new generation of smart, adaptive microscopes that learn to treat the sample as a part of the optical path. *Biobeam* is available as open-source, hardware-agnostic python package at https://maweigert.github.io/biobeam.

**Acknowledgments:** We thank Benjamin Judkewitz (Charité Berlin), Pavel Tomancak, Kaushikaram Subramanian, Loic Royer, and Bevan Cheeseman (all MPI-CBG) for helpful discussions and insightful comments. The authors acknowledge funding from the Max Planck Society.

**Competing Financial Interests:** The authors declare no competing financial interests.

**Correspondence:**  kreysing@mpi-cbg.de